\documentclass [twocolumn, aps, prb, showpacs] {revtex4}
\usepackage{graphicx}
\input{epsf}
\begin{document}

\title {Spin solid phases of spin $1$ and spin $3/2$ antiferromagnets
on a cubic lattice.}

\author{Karol Gregor and Olexei I. Motrunich}
\affiliation{Department of Physics, California Institute of Technology,
Pasadena, CA 91125}
\date{\today}

\begin{abstract}
We study spin $S=1$ and $S=3/2$ Heisenberg antiferromagnets on a
cubic lattice focusing on spin solid states. Using Schwinger boson
formulation for spins, we start in a $U(1)$ spin liquid phase
proximate to Neel phase and explore possible confining paramagnetic
phases as we transition away from the spin liquid by the process of
monopole condensation. Electromagnetic duality is used to rewrite
the theory in terms of monopoles. For spin $1$ we find several
candidate phases of which the most natural one is a phase with spins
organized into parallel Haldane chains. For spin $3/2$ we find that
the most natural phase has spins organized into parallel ladders. As
a by-product, we also write a Landau theory of the ordering in two
special classical frustrated XY models on the cubic lattice, one of
which is the fully frustrated XY model. In a particular limit our
approach maps to a dimer model with $2S$ dimers coming out of every
site, and we find the same spin solid phases in this regime as well.
\end{abstract}

\maketitle

\section{Introduction}

A simple, nontrivial, and physically common example of a regular
system of quantum objects is a collection of spins on a lattice.
This is easiest to analyze if the interactions do not compete and
all prefer the same spin state; the resulting phases have been known
for a long time and include ferromagnetic and Neel states.  A much
richer situation of current interest is when interactions compete.
The frustration together with quantum fluctuations can destroy the
magnetic order and produce spin solid or spin liquid phases. In a
spin solid, spins combine into larger singlet objects such as
valence bonds which form an ordered pattern on a lattice. Such
phases have been found in nature,\cite{Taniguchi, Chow, Kageyama}
and also in numerical studies of model Hamiltonians.\cite{Sandvik,
Beach, Harada} A spin liquid, on the other hand, is a featureless
paramagnet, which can be crudely viewed as a quantum superposition
of many valence bond configurations, thus the name ``resonating
valence bonds'' (RVB) state. So far there are only few experimental
candidates, but on the theoretical side the existence of spin
liquids in many varieties and our understanding of them is well
established (see Ref.~\onlinecite{Banerjee} for a recent collection
of references and also a very recent example of the so-called
Coulomb phase in 3d, which is the spin liquid relevant to the
present work).

In this paper we look for natural spin solid phases of spin $1$ and
spin $3/2$ on a cubic lattice.  A direct study of spin Hamiltonians
that can stabilize such phases is difficult but can be done in some
cases with Quantum Monte Carlo.  Which phases are realized will of
course depend on the specific model:  For example,
Refs.~\onlinecite{Sandvik, Beach} found valence bond solids in spin 1/2
systems with ring exchanges on the square and cubic lattices.
Refs.~\onlinecite{Harada,Harada3d} found spin solid phases for a spin 1
model with biquadratic interaction on the anisotropic square lattice,
but only magnetically ordered phases on the isotropic square and cubic
lattices.

Here we follow instead a more phenomenological approach.
\cite{Arovas, ReadSachdev, Haldane, MotrunichSenthil} A systematic
and commonly used route to achieve this, and the one we start with,
is to generalize the spins to a representation of higher symmetry
group, here taken to be $SU(N)$.\cite{Arovas, ReadSachdev} The
problem can be solved exactly in the $N \to \infty$ limit and one
can consider fluctuations around this limit to get long distance
properties of the system.  This approach, while difficult to connect
with the actual microscopic $SU(2)$ spin system, nevertheless gives
us some guidance about what phases to expect and gives us a form of
the effective field theory. Here it results in a gauge theory which
naturally exhibits deconfined (liquid) and confined (solid) phases,
and we expect that if a microscopic spin system has such phases,
they should be described by this theory.

\begin{figure}[h]
\epsfxsize=3in \centerline{\epsfbox{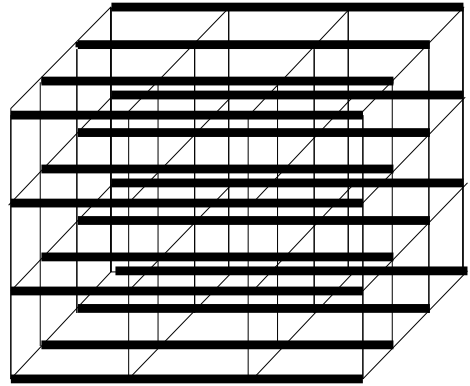}} \caption{The
most natural spin solid phase for $S=1$ on the cubic lattice. The
thick lines denote links with large spin-spin correlations
suggesting that the spins organize into Haldane chains along one
lattice direction.} \label{fig_phase1of1}
\end{figure}

\begin{figure}[h]
\epsfxsize=3in \centerline{\epsfbox{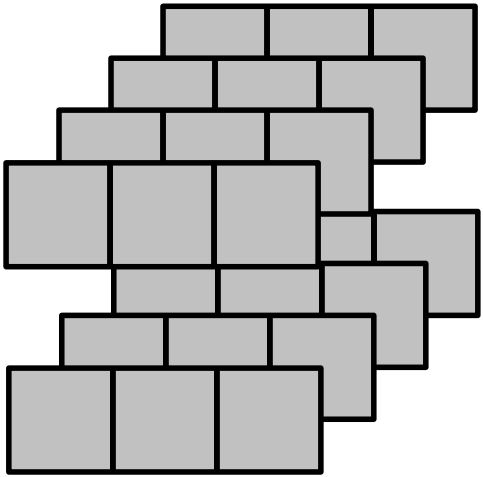}} \caption{The
most natural spin solid phase for $S=3/2$ on the cubic lattice. The
drawn bold lines denote links with large spin-spin correlations
suggesting that the spins organize into ladders.}
\label{fig_phase1of32}
\end{figure}

One of the spin liquid phases expected in 3d is the so-called
Coulomb phase.  It is a compact $U(1)$ gauge theory coupled to
matter in the deconfined phase, where the matter fields (spinons)
are gapped, gauge field (emergent photon) is gapless, and monopoles
(which arise due to compactness) are gapped.  In addition,
importantly, there are spin Berry phases that lead to the presence
of a background charge in the gauge theory formulation.  This makes
the confined phases nontrivial in that they break lattice symmetries
and therefore correspond to various spin solids.  The transition
occurs because the monopoles condense, and the theory can be
equivalently analyzed in terms of them by employing standard
electro-magnetic duality. The background charge causes monopoles to
acquire a phase when they hop around a
plaquette.\cite{MotrunichSenthil} This leads to a nontrivial
monopole condensation pattern, which then corresponds to a spin
solid phase. In 2d the physics is similar, except that the monopoles
are instantons and they always proliferate, so there is no Coulomb
spin liquid. This approach was first used by Read and
Sachdev\cite{ReadSachdev} on the square lattice. The spin solids for
spin 1/2 on the cubic lattice were analyzed in
Ref.~\onlinecite{MotrunichSenthil} and near several different
Coulomb spin liquids in Ref.~\onlinecite{Bernier}.
Ref.~\onlinecite{Bergman} was led (in a different context) to a
gauge theory with background charges on a diamond lattice which was
attacked using analogous techniques.

For the spins on the cubic lattice, the analysis depends only on the
spin magnitude.  Any case can be mapped onto $S=0$, $1/2$, and $1$
in 2d and $S=0$, $1/2$, $1$, and $3/2$ in 3d.  Only the spin 1/2
case was considered so far, but these results cannot be transferred
to the other spins since each requires a separate analysis. This is
the task of the present work. We find that the most natural phases
for spin $1$ and $3/2$ are the ones shown on
Figures~\ref{fig_phase1of1}~and~\ref{fig_phase1of32}. In the $S=1$
case the spins organize into Haldane chains.  This is easiest to
understand in the standard picture where we break spin $1$ into two
spin $1/2$'s and form singlets with spin $1/2$'s of spins on either
side. Similarly, in the $S=3/2$ case we break spin $3/2$ into three
spin $1/2$'s and form singlets on the bonds of the ladders. Several
approaches that we have taken and used in different parameter
regimes suggest the same spin solid states, which gives us
confidence that these phases are very natural in the two cases.

\section{Schwinger bosons, dual reformulation, and a basic phase diagram}
\label{sec:formulation}

\subsection{Schwinger bosons}
We begin by briefly reviewing the standard technique of large
$N$ for spins.\cite{ReadSachdev, Arovas}
This maps (approximately) our spin system into a theory of spinons
coupled to a $U(1)$ gauge field in the presence of static background
charges.  Our main work is the analysis of this theory, while the
purpose of the review here is to establish the connection with the
properties of the original spin system.

The basic steps in the derivation are as follows.
We generalize the $SU(2)$ spin to $SU(N)$ spin and denote it by
$S^{\beta}_{\alpha}(i)$. We write the spins in terms of Schwinger
bosons:
\begin{eqnarray}
S^{\beta}_{\alpha}(i) &=& b^\dagger_{\alpha} (i) b^{\beta}(i)
~~ \mbox{sublattice A} ~, \nonumber \\
S^{\beta}_{\alpha}(j) &=& -\bar{b}^{\beta\dagger} (j) \bar{b}_{\alpha}(j)
~~ \mbox{ sublattice B} ~,
\end{eqnarray}
where the $b,\bar{b}$'s are bosonic operators that transform under
the fundamental representation of $SU(N)$ if the index is on the top
and under its conjugate if the index is on the bottom. To get the
Hilbert space of the spins we need to restrict the boson occupations as
\begin{eqnarray}
b^\dagger_{\alpha}(i) b^{\alpha}(i) &=& n_c ~,\nonumber \\
\bar{b}^{\alpha\dagger} (j) \bar{b}_{\alpha}(j) &=& n_c ~,
\label{eq_Length_Constraint}
\end{eqnarray}
where $n_c$ corresponds to the spin length.
The $SU(N)$ spin Hamiltonian is
\begin{equation}
H = \frac{J}{N} \sum_{\langle i, j\rangle}
S^{\beta}_{\alpha}(i) S^{\alpha}_{\beta}(j) ~,
\end{equation}
which reduces to the $SU(2)$ Heisenberg spin model for spin $S$
when $N=2$ and $n_c = 2S$.

Next we write the system in the path integral picture, imposing the
constraints (\ref{eq_Length_Constraint}) by Lagrange multipliers. The
spin interaction contains quartic terms; to get action that is quadratic
in the boson fields, we use Hubbard-Stratonovich transformation and
obtain
\begin{eqnarray}
{\cal L} &=& \sum_{i}
b^\dagger_{\alpha} (i) \left( \frac{\partial}{\partial \tau}
+ i \lambda(i) \right) b^{\alpha}(i) - i \lambda(i) n_c \nonumber \\
&&+ \sum_{j} \bar{b}^{\alpha\dagger} (j) \left( \frac{\partial}{\partial
\tau} + i \lambda(j) \right) \bar{b}_{\alpha}(j) - i \lambda(j) n_c
\nonumber \\
&&+ \sum_{\langle i, j \rangle} \frac{N}{J} |Q_{ij}|^2 -
Q_{ij}^* b^{\alpha}(i) \bar{b}_{\alpha} (j) + h.c. \label{eq_bQ}
\end{eqnarray}
The path integral goes over $b, \bar{b}, Q, \lambda$.

We can now integrate out the $b$'s. The resulting expression will
have coefficient $N$ in front of it. At large $N$ it can be
approximated by its saddle point value. Our departing point is such
a ``mean field'' with uniform $Q_{r, r+\hat{m}}(\tau) = \bar{Q}$ and
$\lambda(r,\tau) = \bar{\lambda}$ and assuming gapped $b$ spectrum;
this represents a Coulomb spin liquid, which is a stable phase in
three dimensions.  The effective theory is obtained by considering
the fluctuations of the fields, $Q_{r, r+\hat{m}}(\tau) = [\bar{Q} +
q_m(r,\tau)] e^{i \alpha_m(r,\tau)}$ and $\lambda(r,\tau) =
\bar{\lambda} + i \alpha_0(r,\tau)$. Here $r$ runs over all sites of
the cubic lattice and $\hat{m} = \hat{x}, \hat{y}, \hat{z}$ denotes
one of the directions in 3d.  The amplitude fields $q_m$ are
massive, and so are the fields $\alpha_m$ and $\alpha_0$ near the
wavevector $(0,0,0)$. On the other hand, the fields $\alpha_m$ and
$\alpha_0$ near the wavevector $(\pi,\pi,\pi)$ are massless and
describe the gauge field (photon) of the Coulomb phase, $a_m \equiv
\alpha^{(\pi,\pi,\pi)}_m$, $a_\tau \equiv \alpha_0^{(\pi,\pi,\pi)}$.
For details of the derivation, see the original
Ref.~\onlinecite{ReadSachdev} (our notation is slightly different
compared to these papers, which use a two-site unit cell labeling
instead).

As emphasized in Refs.~\onlinecite{ReadSachdev,Haldane},
we also have to consider effect of Berry phases, which is crucial
for the understanding of the spin solid states.
A very convenient encapsulation of the low-energy degrees of freedom
and the Berry phases is provided by the following re-latticized
Euclidean action:\cite{SachdevJalabert, SachdevPark}
\begin{eqnarray}
\label{eq_Sorig}
Z &=& \int_{-\pi}^{\pi} Da_{i\mu} e^{-S_a - S_B} ~, \\
S_a &=& -\beta \sum_{i, \mu < \nu} \cos(\nabla_\mu a_{\nu} -
\nabla_\nu a_{\mu} ) ~, \nonumber \\
S_B &=& i \sum_i \eta_i a_{i \tau} ~. \nonumber
\end{eqnarray}
Here we have a compact $U(1)$ gauge field $a$ residing on the links of
a (3+1)d space-time lattice and described by the action term $S_a$.
The $S_B$ term comes from detailed consideration of the Berry phases,
and $\eta_i$ is $2S$ on one sublattice of the spatial lattice and
$-2S$ on the other one.  In the Hamiltonian language this has a simple
interpretation as a background charge of value $2S$ on one sublattice
and $-2S$ on the other one:
\begin{eqnarray}
\label{Hdimer}
&& H = u \sum_{r,m} E_{rm}^2
- \kappa \sum_{r, m<n} \cos(\nabla_m a_n - \nabla_n a_m) ~,
\\
&& (\nabla \cdot E)_r = \eta_r = \pm 2S ~,
\end{eqnarray}
where $E_m$ are electric fields residing on the links of the 3d cubic
lattice and conjugate to $a_m$.  Thus we obtained a compact $U(1)$ gauge
theory in the presence of background charge.\cite{ZhengSachdev, FradkinKivelson, FradkinBook}

Throughout, we will assume the spinons are gapped and are integrated out.
Note that even though we start in the Coulomb phase where the gauge
field is deconfined, the above action also provides access to confining
paramagnetic phases, and this will be our main focus.
To sum up, we will be describing spin solid phases that are proximate to
the simple Coulomb phase; the latter with the specified Berry
phases encoded in the staggered background charge is in turn appropriate
in the vicinity of the conventional Neel phase.

Since we will continue with (\ref{eq_Sorig}) we need a way to
connect the variables there to the original spin variables. This is
done as follows. The nearest neighbor spin-spin correlation $\langle
{\bf S}_r \cdot {\bf S}_{r'} \rangle$ is proportional to the bond
variable $|Q_{rr'}|^2$. To get the connection between the fluctuation
of the magnitude of $Q$ and the gauge fields we have to also keep the
massive amplitude fields $q_m$ in the above derivation when
integrating out the $b$'s.  One finds that the $(\pi,\pi,\pi)$
component couples to the gauge fields in the action as follows:
$\delta S = i \gamma \sum q^{(\pi,\pi,\pi)}_m (\partial_m a_\tau -
\partial_\tau a_m)$, with some coupling parameter $\gamma$. On the
other hand, in the derivation of the path integral from the
Hamiltonian formulation of the gauge theory, the electric field
is coupled to the gauge field in the same way, i.e., via a term $i
\sum E_m (\partial_m a_\tau - \partial_\tau a_m)$ in the action.
Thus the electric field gives the fluctuation of the staggered
nearest neighbor spin-spin correlation function.

\subsection{Electro-magnetic duality}

\begin{figure}
\epsfxsize=3in \centerline{\epsfbox{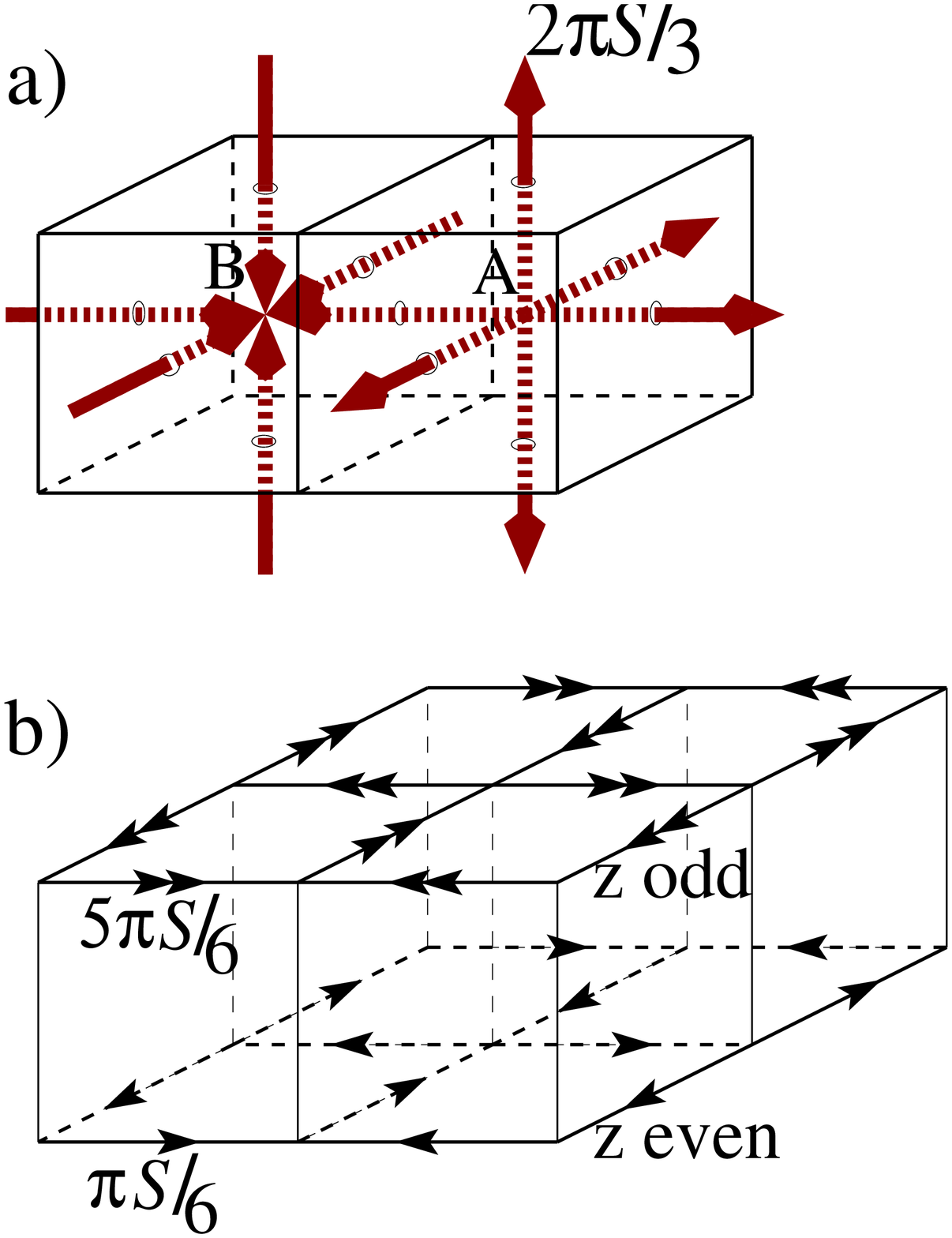}}
\caption{a) Original background electric charges $2S$ and $-2S$ on the
two sublattices give rise to the static dual magnetic fluxes as seen
by the monopoles.  b) Gauge choice for $L^0$ that realizes these fluxes
modulo $2\pi$.
}
\label{fig:gauge}
\end{figure}

We now proceed to the analysis of the model (\ref{eq_Sorig}).
We are interested in the confining phases,
which will necessarily break lattice symmetries for spins $S=1/2, 1,
3/2$ studied here. The confinement occurs due to condensation of
monopoles. Therefore we would like to express the theory in terms of
them. This can be done by the standard electro-magnetic duality. The
duality maps the theory of a compact $U(1)$ gauge field without
charges into a theory of a noncompact gauge field coupled to charges
-- the monopoles of the original theory. The noncompactness comes
from the fact that we have dropped the electric charges in the
original theory; had we retained them, we would have obtained a
compact dual gauge field whose monopoles would correspond to the
original charges. The new variables reside on the lattice dual to
the original lattice. The background charge of the original theory
gives rise to a static dual magnetic flux emanating out of the center
of each cube as drawn in Figure~\ref{fig:gauge}.
This flux alternates in sign from one cube to the next
and frustrates the monopole hopping. Therefore we obtain a theory of
monopoles with frustrated hopping that are coupled to the dual
noncompact gauge field.\cite{MotrunichSenthil}
The duality can be done explicitly with various approximations clearly
displayed as is written in Appendix~\ref{app:duality}.

Explicitly, the partition function is
\begin{eqnarray}
Z &\sim& \int_{-\pi}^{\pi} D\theta \int_{-\infty}^\infty DL \,\,
e^{-S_{\rm dual}} ~, \\
S_{\rm dual} &=& \sum \frac{(\partial L)^2}{8 \pi^2 \beta}
- \sum \lambda \cos(L + L^0 - \nabla \theta) ~,
\label{Sdual}
\end{eqnarray}
where $L$ is the dual gauge field,
$(\partial L)_{\mu \nu} = \nabla_\mu L_\nu - \nabla_\nu L_\mu$
is the four dimensional curl,
$L^0$ is the frustration that results from the original background
charge and ultimately from Berry phases, and $\theta$ is the monopole
field.  A convenient choice of $L^0$ that produces the appropriate
static fluxes is shown in Figure~\ref{fig:gauge}; all subsequent work
is done in this gauge.

\begin{figure}[h]
\epsfxsize=3in \centerline{\epsfbox{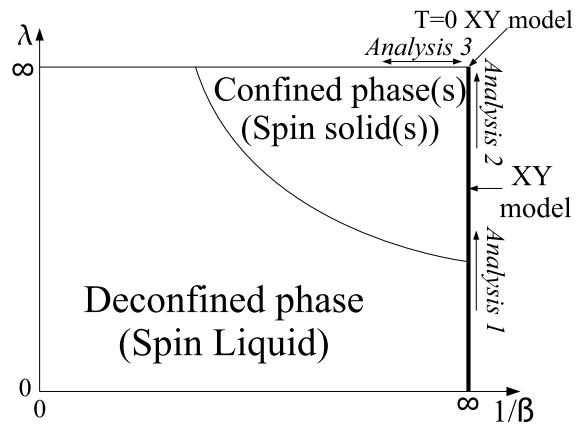}}
\caption{Sketch of the expected phase diagram for the
dual action Eq.~(\ref{Sdual}).
}
\label{gt_phase_diag}
\end{figure}

The advantage of the dual formulation is that it has no sign problem and
can be in principle studied by Monte Carlo. A sketch of the phase
diagram is on Figure \ref{gt_phase_diag}. In the bottom
left side of the diagram, the monopoles are gapped and the system is
in the deconfined phase, which correspond to the Coulomb spin liquid
in the spin model. At large enough $\lambda$ and $1/\beta$, the
monopoles condense. They can condense in various patterns which
translate to various spin solid phases of the original model.
Duality relates the original field theory (\ref{eq_Sorig}) to the
large $\lambda$ part of the dual theory. It is hard to analyze the
transition in the large $\lambda$ limit. Instead we look at three
different places in this phase diagram. At $1/\beta = \infty$ the
system becomes frustrated XY model. First we analyze the phase
transition looking for ordering of the XY spins as we cross the
phase boundary to the ordered phase. This gives us the most likely
monopole condensation patterns near the transition. Next we look at
the classical ground state of the XY model in the upper right corner
of the phase diagram as approached from $1/\beta = \infty$. Finally
we look near the same point but in the limit $\lambda \gg 1/\beta$.

\section{Analysis 1,2: frustrated XY model at $1/\beta = \infty$}
\label{sec:An12}

\subsection{Outline of the Analysis}
\label{subsec:An12}

In this section we describe in general terms the analysis in the
$1/\beta = \infty$ limit where the dual action Eq.~(\ref{Sdual}) reduces
to a frustrated XY model. We look at the phase transition
and the classical ground state.

In the Analysis 1, we consider the transition in the spirit of the
Landau theory.  We identify the relevant low-energy fields, write the
most general quartic potential consistent with the symmetries, and
study it in mean field.
The approach is the same for each spin $S$, but the details are
unique in each case and are contained in
Subsections~\ref{subsec:results1} and \ref{subsec:results32}
for spin 1 and spin 3/2 respectively (spin 1/2 was considered
using this approach in Ref.~\onlinecite{MotrunichSenthil}).

More explicitly, the mean field derivation is done as follows. The
mean field theory of XY spins is described by the continuous soft
spin action
\begin{equation}
\int d\tau \left[
\sum_R |\partial_\tau \Phi_R|^2 -
\sum_{\langle RR' \rangle} (t_{RR'} \Phi_R^* \Phi_{R'} + c.c.) +
\sum_R V(|\phi_R|^2)
\right]
\label{eq_XY_cont}
\end{equation}
with some potential $V(|\Phi_R|^2) = r_0 |\Phi_R|^2 + u_0 |\Phi_R|^4
+ \cdots$.  After crossing the transition from the disordered side, the
system enters a phase with non-zero $\Phi_R$ that minimize this action.
The initial step is to minimize the kinetic energy. This will turn out
to have a three-dimensional manifold of minima for spin 1 and
four-dimensional one for spin 3/2. One then expands around these minima
and writes all terms to a given order that are allowed by symmetry.
In both cases the degeneracy is lifted at the fourth order.
We will find that for spin 1 there are three independent quartic terms
and we manage to draw a general phase diagram of the Landau theory.
For spin 3/2, there are five such terms and the parameter space is too
rich for us to describe the phase diagram completely.  In this case we
confine ourselves to examining the potential obtained from the most
natural microscopic fourth order term and determining its mean field
phase.

In the Analysis 2, we find the classical ground state of the
frustrated XY model -- the state in the upper right point of the
phase diagram Figure~\ref{gt_phase_diag} -- by a direct minimization
of the hard-spin action (\ref{Sdual}). We use the following method:
For some system size, we start with a random configuration of spins.
We pick a random spin and minimize its (local) energy and repeat
this process until the total energy converges. Different starting
configurations will lead to different final energies, because
sometimes the system gets stuck in some local minima. We repeat this
procedure for many starting configurations and also for different
system sizes. We then select the configurations with the same lowest
energy, which gives the absolute minimum of the potential. The case
of spin 3/2, which corresponds to a fully frustrated XY model, was
already considered some time ago in Ref.~\onlinecite{Diep}, and our
method produces results in agreement with that work.

For both spin 1 and spin 3/2, we find that the classical ground state
coincides with the most natural state identified in the mean field
theory near the transition.  This suggests that there is only one
XY-ordered phase along the $1/\beta=\infty$ line in
Fig.~\ref{gt_phase_diag}, which could in principle be tested in
Monte Carlo studies of the corresponding frustrated XY models.

We have described how to find the phases of the dual action in the
$1/\beta = \infty$ limit.  However we are interested in the phases
of the original spin model. To make the connection we calculate
the energies and staggered curls of the monopole currents in the dual
model and relate them to variables in the original spin problem. These
variables are the plaquette energy and the bond expectation value
respectively. This allows us to determine the spin solid patterns.

The mapping of the first variable, the energy, is simple. Energy
simply maps to energy. In the dual model we can calculate the energy
for each bond, which is $\epsilon = 2{\rm Re}(t_{RR'}\Phi^*_R
\Phi_{R'})$. The center of a bond of the dual lattice coincides with
the center of a plaquette of the original lattice, and so the
calculated energy is the plaquette energy of the original model.

The connection of the staggered curls of the monopole current to the
original bond variables is established as follows. The monopole
current is given by $J_M = 2{\rm Im}(t_{RR'}\Phi^*_R \Phi_{R'})$. In
terms of the original gauge theory, Eq.~(\ref{eq_Sorig}), just as
the electric current produces magnetic field, the magnetic current
produces electric field. The resulting electric field is given by
the analog of Biot-Savart law. However, approximately, if we have a
loop of the magnetic current, the electric field it produces in the
center is proportional to the circulation of the current, which is
what we call the curl of the monopole current. As we described in
the preceding section, the electric field is proportional to the
staggered fluctuation of the nearest neighbor spin-spin correlation
function, therefore the claimed connection. We will use this
extensively in the detailed treatment of spin 1 and spin 3/2 below.

\subsection{Results: Spin 1}
\label{subsec:results1}

\subsubsection{Analysis 1: Phase transition of the XY model}

Now we turn to finding the phases for spin 1.
We choose the gauge shown on Figure~\ref{fig:gauge}.
In this case the hopping amplitudes in (\ref{eq_XY_cont}) are given by
\begin{eqnarray}
t_{R,R+\hat{x}} &=& \frac{1}{2}\left[(-1)^z \sqrt{3} + i(-1)^{x+y}\right]
~,\\
t_{R,R+\hat{y}} &=& \frac{1}{2}\left[(-1)^z \sqrt{3} - i(-1)^{x+y}\right]
~,\\
t_{R,R+\hat{z}} &=& 1 ~.
\end{eqnarray}
The band structure has three minima and hence the space of ground
states of the kinetic energy is three dimensional. Convenient choice
of the basis is the following:
\begin{eqnarray*}
\Psi_1 &=& \frac{(-1)^{x+y+z} - (-1)^{x+y} \sqrt{3} + i
[(-1)^z+\sqrt{3}]}{2\sqrt{2}} ~, \\
\Psi_2 &=& i\frac{(-1)^{x+y+z} - (-1)^{x+y} \sqrt{3} - i
[(-1)^z+\sqrt{3}]}{2\sqrt{2}} ~, \\
\Psi_3 &=& - \, \frac{(-1)^y+i(-1)^x}{\sqrt{2}} ~.
\end{eqnarray*}
A general kinetic energy ground state can be written as
\begin{equation}
\Phi(R) = \phi_1 \Psi_1(R) + \phi_2 \Psi_2(R) + \phi_3 \Psi_3(R)
\end{equation}
with complex fields $\phi_{1,2,3}$.
This degeneracy will be lifted by nonlinear terms. To find out how,
we would like to write the Landau theory for the $\phi$'s,
including all terms that are allowed by symmetry. Thus we
need to find how the $\phi$'s transform under the lattice symmetries.

The generators of the symmetries are the translations by one lattice
spacing in the x,y,z directions, 90 degree rotations around the
x,y,z axes (it suffices to consider two out of three rotations), and
mirror reflections. Note that the fluxes seen by the monopoles (and
encoded in the complex phases of the hopping amplitudes $t_{RR'}$)
change sign under unit translations. The original spin problem is
translationally invariant, and this is represented in the dual
action (\ref{eq_XY_cont}) as follows. The fluxes remain unchanged if
the $t$'s are also conjugated after the translation, and there is a
gauge transformation that brings such modified $t$'s to the original
themselves. The action of the symmetry on the field $\Phi$ is then a
combined application of the translation of the coordinates,
conjugation, and gauge transformation. Similar considerations apply
for the 90 degree rotations performed here about the dual lattice
axes. After carrying through this analysis, the transformation
properties of $\phi$'s are remarkably simple:
\newline
\begin{tabular}{|r||r|r|r|r|r|r|}
\hline & $T_x$ & $T_y$ & $T_z$ & $R_x$ & $R_y$ & $R_z$ \\
\hline \hline $\phi_1 \to$ & -- & + & + & + & $\phi_3^*$ & $\phi_2^*$ \\
\hline $\phi_2 \to$ & + & -- & + & $\phi_3^*$ & + & $\phi_1^*$ \\
\hline $\phi_3 \to$ & + & + & -- & $\phi_2^*$ & $\phi_1^*$ & + \\
\hline
\end{tabular}
\newline
\newline
\newline
In this table, ``$+$'' or ``$-$'' stands for $\phi_i \to \phi_i^*$
or $\phi_i \to -\phi_i^*$ respectively. We see that $\phi_1$ can be
loosely associated with the $x$ direction, $\phi_2$ with $y$, and
$\phi_3$ with $z$. We should also point out that under mirror
symmetries in the dual lattice planes the fields transform simply
$\phi_i \to \phi_i$.

There is only one invariant term at the quadratic level:
\begin{equation}
V^{(2)} = m (|\phi_1|^2 + |\phi_2|^2 + |\phi_3|^2) ~,
\label{eq_V1_2}
\end{equation}
where $m$ is a constant.
There are three independent allowed terms at the quartic level, and
the most general quartic potential can be written in the form
\begin{eqnarray}
V^{(4)} &=& u (|\phi_1|^2 + |\phi_2|^2 + |\phi_3|^2)^2 \nonumber \\
&+& v(|\phi_1|^4 + |\phi_2|^4 + |\phi_3|^4) \nonumber \\
&+& w(\phi_1^{*2} \phi_2^2 + \phi_2^{*2} \phi_3^2 + \phi_3^{*2}
\phi_1^2 + c.c.) ~,
\label{eq_V1_4}
\end{eqnarray}
where $u$, $v$, $w$ are constants.

To find the phases of the Landau theory, we simply need to minimize
this potential. Before we start describing the phases, however, it is
useful to introduce bilinears of the fields. The reason is that these are
gauge independent objects whereas the form of $\Psi_{1,2,3}$ and hence
the transformation properties of $\phi_{1,2,3}$ are gauge dependent.
We consider the following bilinears:
\begin{eqnarray*}
B_0 &=& |\phi_1|^2 + |\phi_2|^2 + |\phi_3|^2 ~, \\
F_1 &=& \frac{1}{\sqrt{3}} (|\phi_1|^2 + |\phi_2|^2 - 2 |\phi_3|^2) ~, \\
F_2 &=& |\phi_1|^2 - |\phi_2|^2 ~, \\
D_x &=& \phi^*_3 \phi_2 + \phi^*_2 \phi_3 ~, \\
D_y &=& \phi^*_1 \phi_3 + \phi^*_3 \phi_1 ~, \\
D_z &=& \phi^*_2 \phi_1 + \phi^*_1 \phi_2 ~, \\
N_x &=& i(\phi^*_3 \phi_2-\phi^*_2 \phi_3) ~, \\
N_y &=& i(\phi^*_1 \phi_3-\phi^*_3 \phi_1) ~, \\
N_z &=& i(\phi^*_2 \phi_1-\phi^*_1 \phi_2) ~.
\end{eqnarray*}
The $B_0$ and the groups of $F$'s, $D$'s and $N$'s form irreducible
representations of dimensions 1, 2, 3, and 3 respectively.
The transformation properties of these bilinears are displayed in
the following table

\begin{tabular}{|r||r|r|r|c|c|c|}
\hline & $T_x$ & $T_y$ & $T_z$ & $R_x$ & $R_y$ & $R_z$ \\
\hline
\hline $B_0$ & + & + & + & + & + & + \\
\hline
\hline $F_1$ & + & + & + & $-\frac{1}{2} F_1 + \frac{\sqrt{3}}{2} F_2$
& $-\frac{1}{2} F_1 - \frac{\sqrt{3}}{2} F_2$ & + \\
\hline $F_2$ & + & + & + & $\frac{\sqrt{3}}{2} F_1 + \frac{1}{2} F_2$
& $-\frac{\sqrt{3}}{2} F_1 + \frac{1}{2} F_2$ & $-$ \\
\hline
\hline $D_x$ & + & $-$ & $-$ & + & $D_z$ & $D_y$ \\
\hline $D_y$ & $-$ & + & $-$ & $D_z$ & + & $D_x$ \\
\hline $D_z$ & $-$ & $-$ & + & $D_y$ & $D_x$ & + \\
\hline
\hline $N_x$ & $-$ & + & + & + & $N_z$ & $N_y$ \\
\hline $N_y$ & + & $-$ & + & $N_z$ & + & $N_x$ \\
\hline $N_z$ & + & + & $-$ & $N_y$ & $N_x$ & + \\
\hline
\end{tabular}
\newline
\newline
\newline
We should also add that all bilinears transform trivially under
mirror symmetries in the dual lattice planes.

We next calculate the energies and the staggered curls of the
monopole currents, which, as described in Subsec.~\ref{subsec:An12},
are related to the plaquette energies and bond variables of the
original spin problem. To repeat, the energy is given by
$\epsilon_\mu (R) = 2{\rm Re}(t_{R,R+\hat{\mu}} \Phi_R^*
\Phi_{R+\hat{\mu}})$, and the monopole current is given by $J_\mu
(R) = 2{\rm Im}(t_{R,R+\hat{\mu}} \Phi_R^* \Phi_{R+\hat{\mu}})$. The
staggered curl of the monopole current is what the name suggests,
for example, $f_z \equiv (-1)^{x+y+z} [J_x(R) + J_y(R+\hat{x}) -
J_y(R) - J_x(R+\hat{y})]$.

\begin{figure}[h]
\epsfxsize=2.5in \centerline{\epsfbox{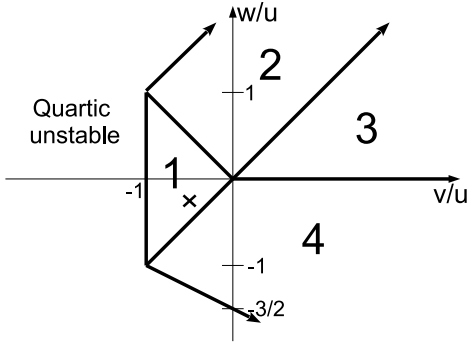}} \caption{Phase
diagram of the Landau theory for spin 1 obtained by minimizing the
potential (\ref{eq_V1_2})+(\ref{eq_V1_4}) for $m<0$ and $u>0$ (the
latter choice is made for concreteness). In the ''Quartic unstable''
region on the left the potential to quartic order is asymptotically
negative and we would have to include sixth order terms to stabilize
it. The cross denotes the parameter point obtained by simply
expanding the microscopic potential $|\Phi|^4$ in terms of the
slowly varying fields $\phi_{1,2,3}$.} \label{fig_ph_diag}
\end{figure}

The energies and staggered curls of the monopole currents are bilinears
in $\phi$ and thus can be expressed in terms of the $B_0, \ldots, N_z$.
They are
\begin{eqnarray}
\epsilon_x &=& \frac{4 B_0}{3} +
\frac{1}{2\sqrt{3}}\left(F_1 + \sqrt{3} F_2 \right)
- 2(-1)^{y+z} D_x \nonumber \\
&+& \sqrt{3} \left[(-1)^y N_y + (-1)^z N_z \right] ~,
\label{eq:Ex_spin1}
\\
f_x &=& 3 \left(F_1 + \sqrt{3} F_2 \right)
- 4(-1)^x N_x ~.
\label{eq:fx_spin1}
\end{eqnarray}
The components in the other directions are obtained from these by
the appropriate rotations using the table, which for all bilinears
except for $F$'s gives the same result as the obvious permutation of
indices. More generally, while the numerical coefficients in these
expressions are obtained from the bare monopole hopping problem, the
overall structure of the contributing terms is dictated by the
symmetries -- one only needs to remember that $\epsilon_x$ and $f_x$
are associated with scalars residing on respectively plaquettes and
bonds of the original spin lattice and also that the rotations and
mirrors quoted here are about the axes and planes passing through
the dual lattice sites. With the above results in hand, we now turn
to analyzing phases of the Landau theory. The phase diagram is
obtained simply by minimizing the potential
(\ref{eq_V1_2})+(\ref{eq_V1_4}) and is shown in
Figure~\ref{fig_ph_diag}. The different phases are described in the
following. In each case the ground state has finite degeneracy; we
display few such states and the others are obtained from them by
obvious permutations; we display nonzero bilinears, the energies,
and the staggered curls of the monopole currents for the first
listed state.

\textbf{Phase 1.} There are three degenerate states. The values in
one of them are
\begin{eqnarray}
&\phi_1 = 1, ~ \phi_2 = \phi_3 = 0; \\
&B_0 = 1; ~ F_1 = \frac{1}{\sqrt{3}}, ~ F_2 = 1; \\
&\epsilon_x = 2, ~ \epsilon_y = \epsilon_z = 1; \\
&f_x = 4\sqrt{3}; ~ f_y = f_z = -2\sqrt{3}.
\end{eqnarray}
The bond variables are drawn on the original spin lattice in
Figure \ref{fig_phase1of1};
they suggest that the spins are organized into Haldane chains along
the x direction. The values of plaquette energies are consistent
with this: the plaquettes in the xy and xz planes are the same and
differ from the plaquettes in the yz plane, $\epsilon_z = \epsilon_y
\neq \epsilon_x$.

\begin{figure}[h]
\epsfxsize=3in \centerline{\epsfbox{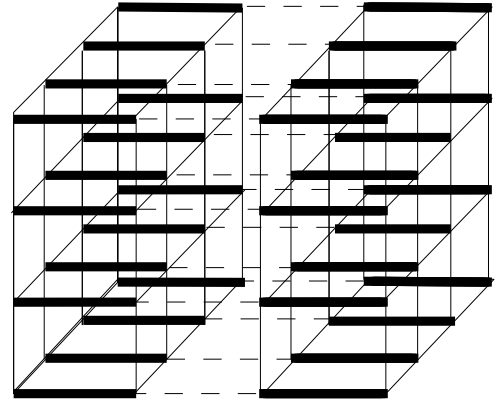}} \caption{Phase
2 of spin 1. The thick lines denote the positions where the bond
variables are strongest and dashed lines where they are weakest.
This suggests that the spins organize into singlets (dimers) and
form a columnar order.} \label{fig_phase2of1}
\end{figure}

\textbf{Phase 2.} There are six degenerate states. The values in one
of them are
\begin{eqnarray}
&\phi_1 = 0, ~ \phi_2 = 1, ~ \phi_3 = \pm i; \\
&B_0 = 2; ~ F_1 = -\frac{1}{\sqrt{3}}, ~ F_2 = -1; ~ N_x = 2; \\
&\epsilon_x = 2, ~ \epsilon_y = \epsilon_z = 3 + 2\sqrt{3}(-1)^x; \\
&f_x = -4[\sqrt{3} + 2(-1)^x], ~ f_y = f_z = 2\sqrt{3}.
\end{eqnarray}
The corresponding drawing of the bond variables on the original spin
lattice is in Figure \ref{fig_phase2of1}, suggesting that
in this phase the spins combine into singlets and form a columnar
dimer state along one direction. Permuting the values of
$\phi_{1,2,3}$ gives six degenerate states that correspond to six
possible ways of placing such columnar solid onto the cubic lattice.

\begin{figure}[h]
\epsfxsize=3in \centerline{\epsfbox{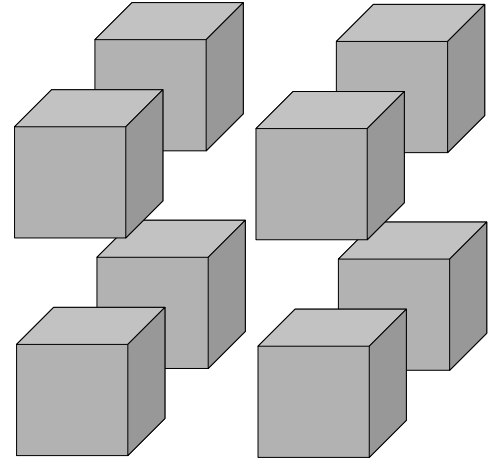}} \caption{Phase 3
of spin 1. The bond variables have higher expectation values on the
cubes shown.} \label{fig_phase4of1}
\end{figure}

\textbf{Phase 3.} There are eight degenerate states specified
as follows:
\begin{eqnarray}
&\phi_1 = 1, ~ \phi_2 = e^{i\alpha_2}, ~ \phi_3 = e^{i\alpha_3}, \\
&\{\alpha_2, \alpha_3\} =
\pm \{2\pi/3, -2\pi/3\}, ~ \pm \{2\pi/3, \pi/3\}, \nonumber \\
&~~~~~~
\pm \{\pi/3, 2\pi/3\}, ~ \pm \{\pi/3, -\pi/3\}; \\
&B_0 = 3; ~ D_x = D_y = D_z = -1; \nonumber \\
&N_x = N_y = N_z = \sqrt{3} \\
&\epsilon_x = 4 + 2 (-1)^{y+z} + 3[(-1)^y + (-1)^z], \text{etc.}, \\
&f_x = -4\sqrt{3} (-1)^x, \text{etc.}
\end{eqnarray}
The nearest neighbor spin-spin correlation has higher expectation
value on the sides of the cubes shown in Figure \ref{fig_phase4of1},
which suggests that this phase corresponds to a box state.
There are eight possible ways of placing such box state onto the cubic
lattice.

\textbf{Phase 4.} There are four degenerate states:
\begin{eqnarray}
&\phi_1 = 1, ~ \phi_2 = e^{i\alpha_2}, ~ \phi_3 = e^{i\alpha_3}; \\
&\alpha_2 = 0,\pi; ~ \alpha_3 = 0,\pi; \\
&B_0 = 3; ~ D_x=D_y=D_z=2;\\
&\epsilon_x = 4 - 4(-1)^{y+z}, ~
 \epsilon_y = 4 - 4(-1)^{z+x}, \nonumber \\
&\epsilon_z = 4 - 4(-1)^{x+y}; \\
&f_x = f_y = f_z = 0 .
\end{eqnarray}
This state breaks lattice symmetries as can be seen from the
plaquette energies.  However, because the bond variables $f_{x,y,z}$
are zero, we do not know a simple interpretation of this phase in
terms of the original spins; some finer characterization than what
we use here is needed to establish this state.

This concludes the discussion of the general phase diagram of the
Landau theory including quadratic and quartic terms.
Higher-order interactions may stabilize some other phases, but the
presented states are the most natural ones.
The actual lowest-energy state depends on the parameters $u, v, w$,
unknown apriori.  If we are to guess which of the four
phases is the most likely candidate in the specific frustrated XY model,
we can consider the simplest microscopic quartic potential $|\Phi|^4$.
When expanded in terms of the continuum fields, we find $u=2$, $v=-1$,
$w=-1/2$; this point is denoted by the cross in Figure \ref{fig_ph_diag}
and lies in the Phase 1, i.e., the Haldane chains phase.


\subsubsection{Analysis 2: The ground state of the XY model}

Minimizing the classical energy of the hard spin XY model as described
in Sec.~\ref{subsec:An12}, we find that the ground state configurations
coincide with the condensate wavefunctions of the phase 1 and hence
the state is that of the phase 1. In particular note that
each wavefunction $\Psi_{1,2,3}$ has the same length $|\Phi|$ on all
sites. The XY angles of spins in this gauge in the three ground states
are
\begin{eqnarray}
&(0,-30,-30,0,60,-90,-90,60) ~, \\
&(0,30,30,0,-60,90,90,-60) ~, \\
&(0,-90,90,180,0,-90,90,180) ~,
\label{eq:angles_spin1}
\end{eqnarray}
where the convention is that we vary position on the cube in the x
direction first, then in the y direction, and then in the z.

\subsubsection{Discussion and extension to anisotropic system}

Some remarks are in order.  First, it is useful to note that the
doublet $F_{1,2}$ can be interpreted as an order parameter of the
Haldane chains phase.  Indeed, one can readily see that the
transformation properties of $F_1$ and $F_2$ coincide with those of
$(Q_x + Q_y - 2Q_z)/\sqrt{3}$ and $Q_x - Q_y$ respectively, where
$Q_m$ is the bond variable in the direction $\hat{m}$. On the other
hand, $N_x$ transforms as $(-1)^x Q_x$ and similarly for $N_y$ and
$N_z$, so $\vec{N}$ can be viewed as an order parameter of the
valence bond solids such as the columnar Phase 2 or the box Phase 3.
In the columnar phase, it is suggestive to view each strong bond in
Fig.~\ref{fig_phase2of1} as representing a singlet formed by two
spin-1's, which can be also drawn as two spin-1/2 valence bonds
connecting the two sites.  However, we should be cautious with such
interpretation, since we can only tell that the deviations of the
bond variables from their mean value will have the displayed
pattern.  The actual state needs to be studied by constructing the
corresponding spin wavefunction.  For example, the Haldane phase of
a spin 1 chain is stable to weak dimerization and should be viewed
as a solid formed by single-strength bonds along the chains, so such
distinct possibilities should be kept in mind.

Let us now assume that the system is in the Phase 1.
It is also interesting to ask what happens when we stretch the
lattice in one of the axis directions, say the z-direction.
In this case the $R_x$ and $R_y$ rotations are no longer symmetries
but the other transformations are. At the quadratic level,
the translation symmetries already prohibit all terms except $B_0$
and $F$'s. Then from $R_z$ we see that only $F_1$ is allowed. Thus
at the quadratic level one new term is allowed. In principle we
should look at the new allowed terms at the quartic level, however we
will assume that this quadratic term is leading but small compared
to the terms that were there before we broke the symmetry.

We find that if the $F_1$ comes with a positive pre-factor, out of
the three ground states it selects the state with chains running
along the z direction whereas if it comes with a negative pre-factor
it selects the two states with chains running along the x and y
directions.

This has a simple interpretation in terms of spins. If the coupling
in the z direction is stronger than in the other directions the
state with maximum number of bonds in this directions is selected
which is the state with chains running in the z-direction. In the
opposite case, the states with fewest bonds in the z direction are
selected which are the states with chains running in the x or y
directions.

\subsection{Results: Spin 3/2}
\label{subsec:results32}

\subsubsection{Analysis 1: Phase transition of the XY model}

We choose the gauge as shown on Figure~\ref{fig:gauge} with $S=3/2$.
The hopping amplitudes are
\begin{eqnarray}
t_x &=& \frac{1}{\sqrt{2}} (-1)^z \left[ 1 + i(-1)^{x+y} \right] ~, \\
t_y &=& \frac{1}{\sqrt{2}} (-1)^z \left[ 1 - i(-1)^{x+y} \right] ~, \\
t_z &=& 1 ~.
\end{eqnarray}
The band structure has four minima and hence the space of the ground
states of kinetic energy is four-dimensional. Unlike the spin 1 case
where this space was three-dimensional and simple basis vectors were
found corresponding to the three directions of the physical space,
there is no such form in the spin 3/2 case. The four wavefunctions
that give us relatively simple subsequent analysis are the following
\begin{eqnarray*}
\Psi_1 &=& (-1)^x \left[\cos\beta - i (-1)^{x+y+z} \sin\beta \right]
~, \label {psi1} \\
\Psi_2 &=& i (-1)^y \left[\cos\beta + i (-1)^{x+y+z} \sin\beta \right]
~, \\
\Psi_3 &=& \frac{1+i(-1)^{x+y}}{\sqrt{2}}
\left[\cos\beta - i (-1)^{x+y+z} \sin\beta \right] ~, \\
\Psi_4 &=& \frac{1-i(-1)^{x+y}}{\sqrt{2}}
\left[\cos\beta + i (-1)^{x+y+z} \sin\beta \right] ~, \label{psi4}
\end{eqnarray*}
where
\begin{eqnarray}
\cos\beta = \sqrt{\frac{\sqrt{3}+1}{2\sqrt{3}}} ~, \quad\quad
\sin\beta = \sqrt{\frac{\sqrt{3}-1}{2\sqrt{3}}} ~.
\end{eqnarray}

We again write $\Phi(R) = \sum_{i=1}^4 \phi_i \Psi_i (R)$.
The transformation properties of the slow fields $\phi_{1,2,3,4}$
are derived in the same manner as in the spin 1 case.
The symmetries are
\begin{eqnarray}
T_x: && \vec{\phi} \to \tau^3 \sigma^0 \vec{\phi^*} ~, \\
T_y: && \vec{\phi} \to \tau^0 \sigma^0 \vec{\phi^*} ~, \\
T_z: && \vec{\phi} \to \tau^1 \sigma^0 \vec{\phi^*} ~, \\
R_y: && \vec{\phi} \to \tau^1 e^{-i \frac{\pi}{4} \tau^2}
     \sigma^1 e^{i \frac{\pi}{3} \sigma^2} \vec{\phi^*} ~, \\
R_z: && \vec{\phi} \to e^{-i \frac{\pi}{4} \tau^3} \sigma^1
     \vec{\phi^*} ~.
\end{eqnarray}
Here $\vec{\phi}$, $\vec{\phi^*}$ are column vectors, and we
introduced two sets of Pauli matrices: $\tau$ matrices that act on the
blocs $\{1,2\}$ and $\{3,4\}$, and $\sigma$ matrices that act within
each bloc ($\tau^0$ and $\sigma^0$ are the corresponding identity
matrices).  At the quadratic order there is one invariant term
\begin{equation}
V^{(2)} = m \sum_{i=1}^4 |\phi_i|^2 ~.
\end{equation}
At the quartic order there are five invariant terms. The expressions
in terms of $\phi$ are rather complicated and not very illuminating,
particularly since $\phi$'s depend on the choice of gauge and the
basis. Instead, we will use gauge invariant bilinears of $\phi$ to
which we now turn.

There are $16$ bilinears and they can be conveniently organized
using tensor product of the introduced two sets of Pauli matrices,
namely $\phi^\dagger \tau^\mu \sigma^\nu \phi$ with
$\mu, \nu = 0,1,2,3$.
These break up into irreducible representations of the cubic lattice
symmetry group.  There are two one-dimensional, one two-dimensional,
and four three-dimensional representations.  The convenient combinations
that we use are
\begin{eqnarray*}
B_0 &=& \phi^\dagger \tau^0 \sigma^0 \phi ~, \\
C &=& \phi^\dagger \tau^0 \sigma^2 \phi ~, \\
F_1 &=& \phi^\dagger \tau^0 \sigma^1 \phi ~, \\
F_2 &=& \phi^\dagger \tau^0 \sigma^3 \phi ~, \\
\vec{D} &=& (D_x, D_y, D_z) = \phi^\dagger \vec{\tau} \sigma^2 \phi ~, \\
\vec{N} &=& (N_x, N_y, N_z) = \phi^\dagger \vec{\tau} \sigma^0 \phi ~, \\
M_x &=& \phi^\dagger \tau^1 (-\frac{1}{2} \sigma^1
                             - \frac{\sqrt{3}}{2} \sigma^3) \phi ~, \\
M_y &=& \phi^\dagger \tau^2 (-\frac{1}{2} \sigma^1
                             + \frac{\sqrt{3}}{2} \sigma^3) \phi ~, \\
M_z &=& \phi^\dagger \tau^3 \sigma^1 \phi ~, \\
K_x &=& \phi^\dagger \tau^1 (\frac{\sqrt{3}}{2} \sigma^1
                             - \frac{1}{2} \sigma^3) \phi ~, \\
K_y &=& \phi^\dagger \tau^2 (-\frac{\sqrt{3}}{2} \sigma^1
                             - \frac{1}{2} \sigma^3) \phi ~, \\
K_z &=& \phi^\dagger \tau^3 \sigma^3 \phi ~.
\end{eqnarray*}
The transformation properties of these bilinears are in the
following table
\newline
\newline
\begin{tabular}{|r||r|r|r|c|c|c|}
\hline
& $T_x$ & $T_y$ & $T_z$ & $R_x$ & $R_y$ & $R_z$ \\
\hline
\hline $B_0$ & + & + & + & + & + & + \\
\hline
\hline $C$ & $-$ & $-$ & $-$ & + & + & + \\
\hline
\hline $F_1$ & + & + & + & $-\frac{1}{2} F_1 + \frac{\sqrt{3}}{2} F_2$
& $-\frac{1}{2} F_1 - \frac{\sqrt{3}}{2} F_2$ & + \\
\hline $F_2$ & + & + & + & $\frac{\sqrt{3}}{2} F_1 + \frac{1}{2} F_2$
& $-\frac{\sqrt{3}}{2} F_1 + \frac{1}{2} F_2$ & $-$ \\
\hline
\hline $D_x$ & + & $-$ & $-$ & + & $D_z$ & $D_y$ \\
\hline $D_y$ & $-$ & + & $-$ & $D_z$ & + & $D_x$ \\
\hline $D_z$ & $-$ & $-$ & + & $D_y$ & $D_x$ & + \\
\hline
\hline $N_x$ & $-$ & + & + & + & $N_z$ & $N_y$ \\
\hline $N_y$ & + & $-$ & + & $N_z$ & + & $N_x$ \\
\hline $N_z$ & + & + & $-$ & $N_y$ & $N_x$ & + \\
\hline
\hline $M_x$ & $-$ & + & + & + & $M_z$ & $M_y$ \\
\hline $M_y$ & + & $-$ & + & $M_z$ & + & $M_x$ \\
\hline $M_z$ & + & + & $-$ & $M_y$ & $M_x$ & + \\
\hline
\hline $K_x$ & $-$ & + & + & $-$ & $-K_z$ & $-K_y$ \\
\hline $K_y$ & + & $-$ & + & $-K_z$ & $-$ & $-K_x$ \\
\hline $K_z$ & + & + & $-$ & $-K_y$ & $-K_x$ & $-$ \\
\hline
\end{tabular}
\newline
\newline
\newline
The energies and staggered curls of monopole currents in term of
these bilinears are
\begin{eqnarray}
\epsilon_x &=& \frac{2}{\sqrt{3}} B_0 - 2 (-1)^{y+z} D_x \nonumber \\
&-& \sqrt{2} \left[(-1)^y M_y + (-1)^z M_z \right] \nonumber \\
&+& \sqrt{\frac{2}{3}} \left[(-1)^y K_y - (-1)^z K_z \right] ~, \\
f_x &=& 2\sqrt{2} (F_1 + \sqrt{3} F_2) + \frac{8 (-1)^x}{\sqrt{3}} N_x ~.
\end{eqnarray}
The components in the other directions are obtained from these by
simple rotations of the coordinates.
Our general discussion following similar
expressions~(\ref{eq:Ex_spin1})~and~(\ref{eq:fx_spin1}) in the
spin 1 case apply here as well (for ease of comparison, we are using
similar labels for objects with identical transformation properties
in the two cases).  However, a word of warning is in order here, which
will be explained in Sec.~\ref{subsubsec:Csym} below.  Observe, for
example, that $\vec{N}$ and $\vec{M}$ have identical transformation
properties and therefore should enter similarly in any expression.
The absence of $M$'s in the expression for $\epsilon_x$ and the
absence of $N$'s in the expression for $f_x$ is due to their different
eigenvalues under an additional artificial symmetry present in the
frustrated XY model, namely a charge conjugation symmetry defined later,
which is also present in our bare kinetic term and thus in the above
expressions.  This symmetry is not physical in the original spin model
and will not be used here; it is therefore important to note that the
degeneracy of the four slow modes obtained from the bare kinetic term
is protected at the quadratic level by the physical lattice symmetries.

There are five independent fourth order terms in $\phi$ allowed by
translation and rotation symmetries:
\begin{eqnarray}
I_1 &=& B_0^2 ~, \\
I_2 &=& C^2 ~, \\
I_3 &=& N_x^2 + N_y^2 + N_z^2 ~, \\
I_4 &=& M_x^2 + M_y^2 + M_z^2 ~, \\
I_5 &=& N_x M_x + N_y M_y + N_z M_z ~.
\end{eqnarray}
As we have said earlier, because the number of invariant terms is large,
we will not attempt to draw the phase diagram of the Landau's theory.
Instead we look at the natural microscopic term
\begin{equation}
V^{(4)} = |\Phi|^4 = \frac{4}{3} I_1 + \frac{1}{3} I_2 - \frac{1}{3} I_3
+ \frac{2}{3} I_4 ~,
\label{V4_spin32}
\end{equation}
where the second equality is obtained after some calculation
keeping only non-oscillatory terms.

This potential does not have any continuous symmetry left other than
the global $U(1)$ transformation of all fields.  In fact the dimensions
of the subgroups of $SU(4)$ that keep the terms $I_1,\ldots,I_5$
invariant are $15,7,6,0,0$ respectively. The potential (\ref{V4_spin32})
achieves global minimum at twelve discrete points.
As an illustration, we consider the following four minima that are
associated with the $z$ direction in the sense to become clear
below:  $(\phi_1, \phi_2, \phi_3, \phi_4) = $
\begin{eqnarray}
(1,0,0,0),~~ (0,1,0,0), ~~ (0,0,1,0), ~~ (0,0,0,1).
\label{gs1-4}
\end{eqnarray}
The four states can be related to each other by translations in the
z direction and rotations about the z axis.
Besides $B_0=1$, the only nonzero bilinears in these states are
$(F_2,N_z,K_z)=(1,1,1)$, $(-1,1,-1)$, $(1,-1,-1)$, and $(-1,-1,1)$
respectively.

The energies are
\begin{eqnarray}
\epsilon_x &=& \frac{2}{\sqrt{3}} \mp (-1)^z \sqrt{\frac{2}{3}} ~, \\
\epsilon_y &=& \frac{2}{\sqrt{3}} \pm (-1)^z \sqrt{\frac{2}{3}} ~, \\
\epsilon_z &=& \frac{2}{\sqrt{3}} ~,
\end{eqnarray}
where the upper sign corresponds to the first and fourth minima and the
lower sign to the other two.

The staggered curls of monopole currents are respectively
\begin{eqnarray}
& f_x, f_y, f_z = \nonumber \\
& 2\sqrt{6}, -2\sqrt{6}, \frac{8 (-1)^z}{\sqrt{3}} ~, \\
& -2\sqrt{6}, 2\sqrt{6}, \frac{8 (-1)^z}{\sqrt{3}} ~, \\
& 2\sqrt{6}, -2\sqrt{6}, -\frac{8 (-1)^z}{\sqrt{3}} ~, \\
& -2\sqrt{6}, 2\sqrt{6}, -\frac{8 (-1)^z}{\sqrt{3}} ~.
\end{eqnarray}
The staggered curls are interpreted as the strength (above some
mean) of the expectation value of nearest neighbor spin-spin
correlation function. The above values imply that the spins organize
themselves into ladders as shown in Figure~\ref{fig_phase1of32},
obtained by drawing say the positive bonds for the first of the
above minima. The four listed states correspond to the four
different positions of ladders with rungs oriented along the z-axis.
The other eight minima are obtained by 90 degree rotations around
the x and y axes and we will not write the specific values of the
variables. The ladder state is natural for $S=3/2$ system, in the
picture where spin $3/2$ breaks up into three spin $1/2$'s and each
of them forms a bond with some other neighboring spin $1/2$.

\subsubsection{Analysis 2: The ground state of the XY model}

We can use the same procedure as in the case of spin 1 to find the
classical ground state of the appropriate XY model.
In fact, this was already done in Ref.~\onlinecite{Diep} because this
problem is the fully frustrated XY model (FFXY), which is of interest by
itself, and we can use the available results. We find that the ground
state configurations coincide with the condensate wavefunctions obtained
above.  Thus, in each of the four displayed states (\ref{gs1-4}),
the microscopic boson field $\Phi$ is given precisely by one of the four
wavefunctions $\Psi_{1, \dots, 4}$.  One can see that $|\Phi|=1$ on all
lattice sites, and the complex phases of $\Phi$ can be interpreted as
angles of the hard-spin XY model.  For example, for $\Phi=\Psi_1$ the
angles are
\begin{equation}
(-\beta, \pi+\beta, \beta, \pi-\beta,
 \beta, \pi-\beta, -\beta, \pi+\beta) ~,
\end{equation}
listed in the same order as in Eq.~(\ref{eq:angles_spin1}). All
other ground states can be obtained by appropriate symmetry
transformations. The agreement of the two analyses suggests that
there is only one ordered phases in the FFXY model, which is also
supported by the available Monte Carlo studies.\cite{Diep,
KimStroud}

\subsubsection{Remark on charge conjugation symmetry in the FFXY}
\label{subsubsec:Csym}
It is worth to point out that the fully frustrated XY model has an
additional charge conjugation symmetry.  Indeed, since $\pi$ and
$-\pi$ fluxes are indistinguishable, $t_{RR'}$ and $t_{RR'}^*$ are
related by a gauge transformation,
$t_{RR'} = e^{i\gamma_R} t_{RR'}^* e^{-i\gamma_{R'}}$,
and so the action remains invariant under the following unitary
 transformation:
\begin{equation}
{\cal C}: \Phi_R \to e^{i\gamma_R} \Phi_R^* ~.
\end{equation}
In terms of the continuum fields, this becomes
\begin{equation}
{\cal C}: \vec{\phi} \to \tau^2 \sigma^2 \vec{\phi^*} ~.
\end{equation}
In particular, the bilinears $N_{x,y,z}$ are odd under ${\cal C}$
while $M_{x,y,z}$ are even, so if this symmetry is included, the $I_5$
quartic term is not allowed (this is why this term did not appear in
Eq.~\ref{V4_spin32} since both the microscopic $|\Phi|^4$ and the bare
quadratic terms in Eq.~\ref{eq_XY_cont} have this additional symmetry).
Thus, the complete field theory for the FFXY model is a $\phi^4$ theory
with four complex fields and independent quartic terms $I_{1, \dots, 4}$.

One consequence of the charge conjugation symmetry is that, for
example, if we draw the $\Psi_1$ state using negative values of the
staggered curls $f_{x,y,z}$ as opposed to using positive values
which was done in Fig.~\ref{fig_phase1of32}, we would obtain another
set of ladders that go perpendicularly to the ones displayed and are
shifted up by one lattice spacing.
To put this in other words, the $\Psi_1$ and $\Psi_4$ states that can be
related by a translation in the z direction followed by a rotation around
the z axis are also related by ${\cal C}$.
In this sense, each of the states Eq.~(\ref{gs1-4}) does not
define a direction in the x-y plane since the correlations in the x and
y directions are related by the charge conjugation symmetry.

Tracing back to the original gauge theory formulation, this symmetry is
present in the simplest model Eq.~(\ref{eq_Sorig}) for $S=3/2$ that we
wrote down and the corresponding simplest ``dimer model'' Hamiltonian
Eq.~(\ref{Hdimer}).  Specifically, the transformation $E \to 1 - E$ on
the links oriented from one sublattice to the other, or equivalently
$1 \leftrightarrow 0$ in the dimer language, takes the model
corresponding to spin $S$ to the one corresponding to spin $3-S$,
while the $S=3/2$ case maps back onto itself.  This symmetry is useful
in the specific models, but there is no corresponding symmetry in the
microscopic derivation from the spin model, and therefore it was not
used in the preceding analysis.

Let us look what happens to the ground states when we add small term
that breaks the charge conjugation symmetry, the $I_5$, to the
potential. Using general arguments it is easy to check that the
twelve minima will shift but not split, and the twelve-fold
degeneracy remains since all are related to each other by lattice
symmetries. Furthermore, each ground state stays translationally
invariant along the ladders and perpendicular to the plane of
ladders (otherwise, if this were not true, there would be more than
twelve states). In other words, the states still have the structure
of ladders. However since the charge conjugation is broken, it is no
longer true that the negative bonds are of the same magnitude as the
corresponding positive ones. This makes sense when interpreted in
terms of spins. In the picture where spin $3/2$ breaks up into three
spin $1/2$ and ladders of valence bonds are formed, the links that
belong to these ladders are different from the links without bonds
(which also form ladders). For example, the system is entangled
along the former but not along the latter. Thus these two should not
be related by any symmetry.

Explicitly, the four states in Eq.~(\ref{gs1-4}) become
\begin{equation}
(1, \delta, 0, 0), ~~ (\delta, 1, 0, 0), ~~ (0, 0, 1, \delta), ~~
(0, 0, \delta, 1),
\end{equation}
with appropriate $\delta$ obtained from minimization.
There is now an additional non-zero bilinear $M_z$, and also
both $F_1$ and $F_2$ are non-zero. The expressions for the energies
and staggered curls in the x and y directions are no longer related,
and we can then associate a unique x or y direction with each of
the four states.  These are ladders with rungs oriented in the
z direction and are related to each other by the z translations and
rotations.

\subsubsection{Extension to anisotropic system}
As in the spin 1 case, we ask what happens when we stretch the system
along one axis, say the z-direction. Again, the $R_x$ and $R_y$
rotations are no longer symmetries but the translations and $R_z$ are.
At the quadratic level, the translation symmetries already prohibit all
terms except $B_0$ and $F$'s. Then from $R_z$ we see that only $F_1$ is
allowed. Thus at the quadratic level one new term is allowed.

We find that if the $F_1$ comes with a positive pre-factor, out of
the twelve ground states it selects four with the ladders that
lie entirely in the x-y plane, whereas if it comes with a negative
pre-factor it selects four states with the ladders running along the
z-direction.  Note that this breaking up into groups of four is a
consequence of the remaining symmetries in the system.

These results have a simple physical interpretation for the spin
system. If the coupling in the z direction is weaker than in the
other directions, the states with fewest bonds in the z direction
are selected which are the states with the ladders lying in the x-y
plane. On the other hand, if the coupling in the z direction is
stronger, the states with the largest number of bonds in the z
direction are selected, which are the states with ladders oriented
in the z direction.

\section{Analysis 3: Mapping to dimers at $\lambda \gg 1/\beta \gg
1$}

Here we look at the right hand corner of the phase diagram
Fig.~\ref{gt_phase_diag} in the regime with $\lambda \gg 1/\beta \gg
1$, where as we will see the system can be mapped to dimers.
\cite{ZhengSachdev,FradkinKivelson,FradkinBook}

The analysis proceeds as follows. First we gauge away the $\nabla
\theta$ in Eq.~(\ref{Sdual}) to obtain
\begin{equation}
S_{\rm dual} = \sum \frac{(\partial L)^2}{8 \pi^2 \beta} - \sum
\lambda \cos(L + L^0) ~, \label{Sdual_L}
\end{equation}
Because we assume $\lambda \gg 1/\beta \gg 1$ the configurations
that contribute significantly to the partition function can be
written in the form $L = -L^0 + 2 \pi n + \delta L$ where $n$ is an
integer and $\delta L$ is small. Note that the $\lambda$ term does
not depend on $n$ and the $1/\beta$ term has a gauge invariance $n
\to n + \nabla m$ where $m$'s are integers on sites. The partition
function can be written as a sum over the gauge equivalent classes.
These classes are in one-to-one correspondence with the fluxes
$j=\partial n$ which are integers on plaquettes, where $\partial n$
is the four dimensional curl $(\partial n)_{\mu \nu} = \partial_\mu
n_\nu -\partial_\nu m_\mu$.

Consider first configurations with $\delta L=0$. Some configurations
of $j$ minimize the action and we denote them by $j^{gs}$. As we
show below, there is an extensive number of them in all our cases. The
configurations with $j$ that are not $j^{gs}$ are at energy of at
least $\sim 1/\beta$ higher. Now turning on $\delta L$, if we show
that the typical energy of excitation in $\delta L$ around a given
$j$ is much smaller then $1/\beta$ then we can neglect all
configurations which are not around $j^{gs}$. We will assume that
this is true and show this self-consistently below.

We define $J^{gs} = -\partial L^0/(2\pi) + j^{gs}$. We expand the
action to the second order and drop the terms that do not depend on
$J^{gs}, \delta L$ to obtain
\begin{eqnarray}
S \approx \sum \frac{4 \pi J^{gs} \cdot (\partial \delta L)
      + (\partial \delta L)^2}
     {8\pi^2\beta}
+ \sum \frac{\lambda}{2} (\delta L)^2 ~. \label{eq_SdeltaL}
\end{eqnarray}
This is just a gaussian integral. There are two quadratic terms and
the first one has $1/\beta$ in front and contains two derivatives
while the second has $\lambda$ in front and contains no derivatives.
Since we are on a lattice the derivatives are of order one. Since
$\lambda \gg \beta$, the first term can be neglected. Next we sum by
parts and integrate out the $\delta L$. Before we do this however,
we notice that the coupling is ferromagnetic in time direction and
$L^0$ has zero time components and its spatial components do not
depend on time. This implies that the $j^{gs}$ and $J^{gs}$ have
zero time components and their spatial components do not depend on
time. Thus we drop time components and time derivatives from the
action and treat the $J^{gs}$ and $L^0$ as three-dimensional. Now we
integrate out the $\delta L$ and obtain
\begin{eqnarray}
S_{\rm eff}[J^{gs}] &=& -\frac{1}{8 \pi^2 \beta^2 \lambda} \sum
(\nabla \times J^{gs})^2 \label{eq_doublecurl}
\end{eqnarray}
Thus, to obtain a ground state, we need to maximize the sum of the
squares of curls of $J^{gs}$.

Let us check the consistency of our approach. From
(\ref{eq_SdeltaL}), $\delta L \sim \nabla \times J^{gs}/(\lambda
\beta)$ and so energy$\sim 1/(\lambda \beta^2)$. This needs to be
much smaller then $1/\beta$ which implies $\lambda \gg 1/\beta$
which is what we assumed.

\begin{figure}[h]
\epsfxsize=3in \centerline{\epsfbox{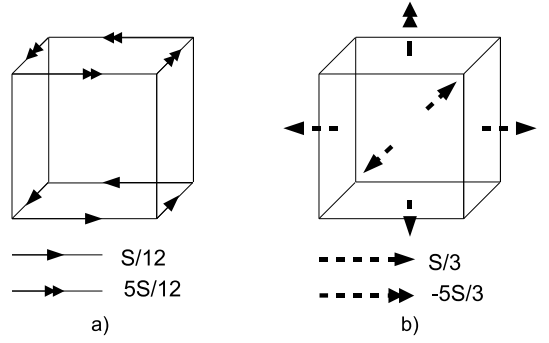}} \caption{ a)
$L^0/(2\pi)$ where $S=1/2, 1, 3/2$ is the spin. The link variables
switch orientations under elementary translation in the x or y
direction. b) The fluxes $(\nabla \times L^0)/(2\pi)$. This figure
is similar to Fig. \ref{fig:gauge} with $2\pi$'s removed to simplify
the discussion of the dimer ground states.} \label{fig_Landcurl}
\end{figure}

Now let us turn to the specific cases of spins. Since the spin 1/2
case has not been considered using this approach before, we will add
it here for completeness. The gauge choice for $L^0$ and the fluxes
$\nabla \times L^0/(2\pi)$ through the faces of the spatial cubes
are shown on Figure \ref{fig_Landcurl} with $S=1/2$. It is easy to
see that the set of ground states consists of all configurations
with precisely one $-5/6$ and five $1/6$ fluxes $J^{gs}$ coming out
of every site of one sublattice of the original spin lattice (and
coming into every site on the other sublattice). $J^{gs} = -\nabla
\times L^0/(2\pi)$ is one such configuration in the spin 1/2 case,
but there are many more.  Associating the $-5/6$ plaquettes with
dimers on the links of the original spin lattice, the set of the
ground states is thus the set of dimer configurations with one dimer
coming out of every site.

Now turn to the case of spin 1. The fluxes $(\nabla \times
L^0)/(2\pi)$ are shown on Figure \ref{fig_Landcurl} with $S=1$.  If
we try $J^{gs}=-\nabla \times L^0/(2\pi)$, each cube contributes
$1/\beta$ energy term proportional to $5 (1/3)^2 + (5/3)^2 = 10/3$.
However we can do better. Using $L = -L^0 + 2\pi n$, if we pick
$n=1$ on the upper link on the front face and zero elsewhere on the
cube in Fig.~\ref{fig_Landcurl}, we lower the magnitude of the flux
on the upper face, at the expense of increasing the flux through the
front face.  The energy of this cube is then $4 (1/3)^2 + 2 (2/3)^2
= 4/3$, which is lower.  It is easy to show that this is the lowest
we can achieve and that the ground state configurations have two
fluxes of value $-2/3$ and four fluxes of value $1/3$ coming out of
every site of one sublattice of the original spin lattice.
Associating the $2/3$ links with dimers, the set of the ground
states is thus the set of dimer configurations with two dimers
coming out of every spin site.

Finally, in the $S=3/2$ case, it is easy to see that the ground
state configurations have precisely three $-1/2$ and three $1/2$
fluxes coming out of every site of one sublattice of the original
cubic lattice. Associating the $-1/2$ links with dimers, the set of
the ground states is thus the set of dimer configurations with three
dimers coming out of every spin site.

Thus, as claimed, in each case there is an extensive number of
$J^{gs}$'s.  To find the true ground state, we need to minimize
(\ref{eq_doublecurl}) among these dimer configurations. It is not
hard to show that for the spin $1/2$ we get columnar state, for spin
$1$ the Haldane chains state of Fig. \ref{fig_phase1of1} and for
spin $3/2$ the ladder state of Fig. \ref{fig_phase1of32}.

Finally we note that defining $E^{gs} = S/3 - J^{gs}$, the set of
$E^{gs}$ is the set of electric fields on links,
cf.~Eq.~(\ref{Hdimer}), with the property that the magnitude of each
is either zero or one (which can be imposed by minimizing the energy
term $\sum E^2$); the mapping between such electric fields and
dimers above is the standard one on the cubic lattice
\cite{ZhengSachdev,FradkinKivelson,FradkinBook}. The final ground
state selection is obtain by maximizing $\sum (\nabla \times E)^2$.


\section{Conclusions}
In this paper we looked for spin solid phases in the system of spin
$1$ and $3/2$ on the cubic lattice. We wrote the spins in terms of
Schwinger bosons, assumed the uniform Coulomb spin liquid phase and by
process of monopole condensation transitioned into spin solid
phases. Using the duality we rewrote the system in terms of monopoles
coupled to a noncompact $U(1)$ gauge field, Eq.~(\ref{Sdual}),
and analyzed this theory in three different limits shown in
Figure~\ref{gt_phase_diag}.

In the first two limits the theory becomes a frustrated XY model.
For spin $1$ the frustrating flux through every plaquette is $2\pi/3$,
while for spin $3/2$ it is $\pi$.  In the first approach, using
symmetries we wrote the Landau's theory near the ordering transition.
It is a $\phi^4$ theory with $\phi$ a complex vector with three
components for $S=1$ and four components for $S=3/2$.  At the quadratic
level only the rotationally invariant mass term is allowed.  At the
quartic level there are three allowed terms for spin $1$ and five for
spin $3/2$.
For spin $1$ we draw a mean field phase diagram Figure \ref{fig_ph_diag}.
For spin $3/2$ we didn't attempt it due to a large number of parameters.
In both cases we also considered the most natural microscopic potential and
found that it selects a state with parallel Haldane chains of
Figure~\ref{fig_phase1of1} for $S=1$ and a state with parallel ladders
of Figure~\ref{fig_phase1of32} for $S=3/2$. These are natural states
for the spin systems to be in, in the picture where spin $1$ breaks up
into two and spin $3/2$ into three spin $1/2$'s and each such spin
$1/2$ forms a singlet bond with another spin $1/2$ of some neighbor.

In the second approach we looked at the classical ground states of the
frustrated XY models and found that these actually describe the same
phases as the most natural ones identified near the transition.

In the third approach the theory becomes a dimer model with $2S$
dimers coming out of every site.  Dimer configurations with
parallel lines for spin $1$ and parallel ladders for spin $3/2$ are
selected, which is the same result as in the other two limits
suggesting that these are indeed the most natural valence bond
solids in the corresponding spin systems. It would be interesting to
look for such spin solid phases in Quantum Monte Carlo studies of
models on the cubic lattice.\cite{Beach, Harada3d}

It is also worth noting\cite{Bergman} that if we consider our quantum
3d systems at a finite temperature, we obtain simply the corresponding
classical 3d dimer models, e.g., with the classical energy given by the
first term in Eq.~(\ref{Hdimer}).  Our results then provide appropriate
long-wavelength (dual) description of the dimer ordering patterns
transitioning out of the so-called Coulomb phase of the classical dimer
models,\cite{Huse, Hermele, Alet} stressing in particular a composite
character of the naive order parameters for the valence bond solid
phases.
It would interesting to explore such 3d classical dimer models and their
transitions further.

\appendix

\section{Classical $U(1)$ duality with background charge}
\label{app:duality} In this section we derive duality for classical
compact $U(1)$ gauge theory.\cite{Peskin,MotrunichSenthil} However
we will use a general notation of antisymmetric tensors, or
differential forms which are fields of antisymmetric tensors. Thus
the derivation will work not only for the gauge theory, whose
objects are one dimensional, but for general $n$-dimensional
objects. For $n=0$ this is the vortex duality of the XY model and
for $n=1$ the duality of the gauge theory. The further advantage of
this derivation is that the formulas are simpler and more
transparent.

First we give the basic notations and properties of antisymmetric
tensors. An $n$-dimensional antisymmetric tensor $\omega$ in $d$
dimensions is a collection of numbers
$\omega_{\mu_1,\mu_2,\ldots,\mu_n}$, where $\mu_v = 1,\ldots,d$,
which is completely antisymmetric. A differential form
$\omega(\vec{r})$ is a field of these tensors.

We define two operations. First is the exterior derivative
$\partial$. The derivative of $\omega$, denoted $\partial \omega$ is
the $(n+1)$-form
\begin{equation}
(\partial \omega)_{\mu_1,\mu_2,\ldots,\mu_{n+1}} = \frac{1}{n!}
\sum_p (-1)^p
\partial_{\mu_{p_1}} \omega_{\mu_{p_2},\ldots,\mu_{p_n}}
\end{equation}
where the sum is over all permutations of the $n+1$ indices and
$(-1)^p$ is $-1$ if the permutation is odd and $1$ if it is even.
Thus for example for $n=1$, a vector field, $(\partial \omega)_{12}
= \partial_1 \omega_2 - \partial_2 \omega_1$ and hence this is the
curl of a vector field.

The second operation that we define is the star operator that takes
$n$-form to $(d-n)$-form
\begin{equation}
(*\omega)_{\nu_1,\ldots,\nu_{d-n}} = \frac{1}{n!}
\epsilon_{\nu_1,\ldots,\nu_{d-n},\mu_{d-n+1},\ldots,\mu_d}
\omega_{\mu_{d-n+1},\ldots,\mu_d}
\end{equation}
where $\epsilon$ is the fully antisymmetric tensor in $d$ dimensions
and repeated indices are summed over. For example in three
dimensions for $n=2$, $(*\omega)_1 = \frac{1}{2}(
\omega_{23}-\omega_{32})$. Note that $**=(-1)^{n(d-n)}$.

A common operator is divergence which in this notation is
proportional to $*\partial*$. As easily checked,
\begin{eqnarray}
(\nabla \cdot \omega)_{\mu_1,\ldots,\mu_{n-1}} &\equiv& \partial_\nu
\omega_{\nu,\mu_1,\ldots,\mu_{n-1}} \\
&=& (-1)^{(n-1)(d-n)} (*\partial* \omega)_{\mu_1,\ldots,\mu_{n-1}}
\end{eqnarray}
For a vector field this is the standard divergence.

We will work on the lattice. The variables are defined on discrete
points. We will define the coordinates of a given variable to be
those of the center of the object the variable belongs to. For
example the $x$ component of a one form $\omega$ in $d=3$ lies on a
link pointing in $x$ direction and it is denoted by
$\omega_x(x+1/2,y,z)$. The $\partial$ now denotes the difference
operator. For example the curl of the $\omega$ is $(\partial
\omega)_{xy}(x+1/2,y+1/2,z) = \omega_y (x+1,y+1/2,z) - \omega_y
(x,y+1/2,z)-\omega_x (x+1/2,y+1,z) + \omega_x(x+1/2,y,z)$.

Finally we will write the integration (summation) by parts
\begin{equation}
\sum \omega \cdot \partial \phi = -\sum (\nabla \cdot \omega) \cdot
\phi + \mbox{surface term}
\end{equation}
where the dot is the sum over the component by component product of
two forms of the same $n$. Note that $*\omega_1 \cdot
*\omega_2 = \omega_1 \cdot \omega_2$. Because we use periodic boundary conditions below, the
surface term will be zero.

Now we are ready to turn to the duality. Let $a$ be an $n$-form in
$d$ dimensions where its variables are defined on the unit circle.
The action is
\begin{equation}
S = -\beta \sum \cos(\partial a) - i \sum \eta \cdot a
\end{equation}
In the first term one takes every component at every point, takes
cosine of it and sums. In the second term the $n$-form $\eta$
denotes the background charge. For the action considered in this
paper, the first term is the $S_a$ and the second term the $S_B$ in
(\ref{eq_Sorig}), while the $\eta$ is the four dimensional vector
with the time component being $\pm 2S$ and the other components
being zero.

The duality proceeds by the following steps.
\begin{eqnarray}
Z &=& \int_{-\pi}^{\pi} Da e^{\beta \sum \cos (\partial a) + i \sum
\eta \cdot a} \nonumber \\
&\approx& \int_{-\pi}^{\pi} Da \sum_p e^{-\frac{\beta}{2} \sum
(\partial a - 2 \pi p)^2 + i \sum \eta \cdot a} \nonumber \\
&=& \int_{-\infty}^{\infty} Da \sum_{q'} e^{-\frac{\beta}{2} \sum
(\partial a - 2 \pi \partial^{-1} q')^2 + i \sum \eta \cdot a} \nonumber \\
&=& \int_{-\infty}^{\infty} Da \sum_{q'} \int_{-\infty}^{\infty}
DJ \nonumber \\
&& \times e^{-\sum \frac{J^2}{2\beta} + i\sum J \cdot (\partial a -
2 \pi
\partial^{-1} q') + i \sum \eta \cdot a} \nonumber \\
&=& \sum_{q'} \int_{-\infty}^{\infty} DJ e^{-\sum \frac{J^2}{2\beta}
- i 2 \pi \sum J \cdot
\partial^{-1} q'} \nonumber \\
&& \times \Delta (\nabla \cdot J - \eta) \label{eq_duality5}
\end{eqnarray}
All numerical factors are dropped throughout, while the sign
``$\approx$'' is used when an approximation is being made that does not
change the qualitative aspects.

In the second line we use the Villain form of the cosine. In the
third line we have written the field $p = \partial \alpha +
\partial^{-1}q'$ as a curl of $\alpha$ plus a field of a particular
monopole current configuration $q'$, $\partial^{-1}q'$. The
$\partial^{-1}$ denotes a particular configuration of $p$ that gives
the monopole currents - that satisfies $q '= \partial p$. Then we
shifted $a \to a - \alpha$. The summation over $\alpha$ extends the
integration of $a$ over the whole real line. The prime on $q'$
denotes that fact that we are summing over fields for which
$\partial q' = 0$.

The third line can be obtained from the fourth one by completing the
square, shifting $J$ and integrating it out.

In the fifth line, the $\Delta$ denotes that the operator inside of
it is zero. This line is obtained from the fourth one by integrating
(summing) by parts and integrating out the $a$.

Next, as shown explicitly below, in our case there are fields $J_0$
and $L_0$ such that
\begin{eqnarray}
\eta &=& \nabla \cdot J_0
\label{eq_eta} \\
J_0 &=& (*P\partial L_0)/2\pi \\
\partial J_0 &=& 0
\label{eq_J_0_0}
\end{eqnarray}
The $P$ shifts a real number by a multiple of $2\pi$ so that the
result is in the interval $(-\pi,\pi ]$.

Using (\ref{eq_eta}) in (\ref{eq_duality5}) we see $\partial *
(J-J_0) = 0$ and hence we can write
\begin{equation}
J = J_0 + (*\partial L)/2\pi
\end{equation}
for some field $L$. To substitute this into (\ref{eq_duality5}) we
notice the following
\begin{eqnarray*}
J^2 = J_0^2 + (* \partial L/2\pi)^2 + 2 J_0 \cdot * \partial L/2\pi
\simeq J_0^2 + (* \partial L/2\pi)^2 .
\end{eqnarray*}
The $\simeq$ denotes that these expressions are equal under integration,
which follows from Eq.~(\ref{eq_J_0_0}).  Also
\begin{eqnarray*}
\partial^{-1} q' \cdot * \partial L
&\simeq& - * \partial \partial^{-1} q' \cdot L
= - *q' \cdot L \equiv -Q' \cdot L , \\
e^{i\partial^{-1} q' \cdot * P\partial L_0}
&=& e^{i\partial^{-1} q' \cdot * \partial L_0}
\simeq e^{-iQ' \cdot L_0} ,
\end{eqnarray*}
where $Q' \equiv *q'$, and we have dropped inconsequential $\pm$ signs;
in the last line, the $P$ can be removed because the resulting
expression, which is in the exponent, differs from the original one by a
multiple of $2\pi$.

With this we can proceed to complete the duality
\begin{eqnarray}
Z &\approx & \sum_{Q'} \int_{-\infty}^{\infty} DL e^{-\sum
\frac{(\partial L)^2}{8 \pi^2 \beta} + i \sum Q' \cdot (L+L_0)} \nonumber \\
& = & \sum_{Q} \int_{-\pi}^{\pi} D\theta \int DL e^{-\sum
\frac{(\partial L)^2}{8
\pi^2 \beta} + i \sum Q \cdot (L+L_0-\partial \theta)} \nonumber \\
&\approx & \sum_p \int_{-\pi}^{\pi} D\theta \int DL e^{-\sum
\frac{(\partial L)^2}{8 \pi^2 \beta} - \frac{\lambda}{2}
(L+L_0-\partial \theta -2 \pi
p)^2} \nonumber \\
& \approx& \int_{-\pi}^{\pi} D\theta \int DL e^{-\sum
\frac{(\partial L)^2}{8 \pi^2 \beta} - \lambda \cos(L+L_0 - \partial
\theta)} \label{eq_Z_dual_5}
\end{eqnarray}
In the first line the summation over $Q'$ is over integer fields
$Q'$ with zero divergence $\nabla \cdot Q'=0$ - currents. In the
second line we introduced $\theta$ that imposes this constraint as a
Lagrange multiplier and summed by parts. In the third line we added
a small term $\sum Q^2/2\lambda$ and assumed that it is not going to
change the basic behavior of the system. Then we summed out $Q$,
which introduced integer $p$ because $Q$ is an integer (this is the
Poisson summation formula). The second term is the Villain form of
cosine. In the last line we approximated it by cosine.

To complete it remains to find $J_0$ and $L_0$. The $\eta$ has values
$\eta_{\tau} (x,y,z,\tau+1/2) = (-1)^{x+y+z} 2S$ and zero for other
components. As easily checked
\begin{equation}
(J_0)_{\tau x} (x+1/2,y,z,\tau+1/2) = \frac{2S}{6} (-1)^{x+y+z}
\end{equation}
and similarly for $y$ and $z$ with other components (other then the
ones obtained by permutation of indices) being zero. This gives the
right $\eta$ and satisfies $\partial J_0=0$. The $L_0$ can be chosen
as on the Fig. \ref{fig:gauge}.

In the final expression (\ref{eq_Z_dual_5}) the $L$ is $1$-form and
hence a gauge field. The $\theta$ is $0$-form - a number on a circle
- a matter field.
Thus we obtained a noncompact $U(1)$ gauge theory coupled to scalar
fields of monopoles with frustrated hopping.

\end{document}